\documentclass[prl,showpacs,showkeys,preprintnumbers,aps,amssymb]{revtex4}
\usepackage[dvips]{graphicx}
\usepackage{amsmath,amssymb}

\begin{document}

\preprint{UT-Komaba/04-7}

\title{Skyrmions coupled with the electromagnetic field via 
the gauged Wess-Zumino term}
 
\author{Munehisa Ohtani}
\email{ohtani@rarfaxp.riken.jp}
\affiliation{Radiation Laboratory, RIKEN, Wako, Saitama 351-0198, Japan}
\author{Koichi Ohta}
\affiliation{Institute of Physics, University of Tokyo,
Komaba, Tokyo 153-8902, Japan }
\date{June 16, 2004}

\vspace{1cm}
\begin{abstract}
 In soliton models expressed in terms of 
the non-linear chiral field, 
the electric current has an anomalous gauge-field contribution 
as the baryon current does.
 We study the spin polarized Skyrmions coupled 
with the electromagnetic field via the gauged Wess-Zumino term
and calculate configurations of the Skyrmion and the gauge field 
with boundary conditions to ensure the physical charge number
for baryons. 

Although the electromagnetic field via
the gauged Wess-Zumino term affects physical quantities in small amounts,
we find that the magnetic field forms
a dipole structure owing to a circular electric
current around the spin quantization axis of the soliton.
This is understood on an analogy with the Meissner effect 
in the super conductor.
The electric charge distributions turn out to have characteristic
structures depending on the total charge, which suggests the intrinsic
deformation of baryons due to orbital motions of the constituents.
\end{abstract}
\pacs{11.10.Lm, 12.39.Dc, 13.40.-f, 14.20.Dh}
\keywords{Skyrme model, topological soliton, Wess-Zumino term, 
charge distribution}
\maketitle
\section{Introduction}
Of recent years the proton spin puzzle has attracted much attention.
As reported in Ref.\ \cite{emc}, the quark spin is found to be
responsible for only a small amount of the proton spin.
The gluon spin contribution is also estimated from the global analysis
of the parton distribution function\cite{aac}, but has yet large uncertainties.
The experiment of polarized proton collision at RHIC in progress
is expected to give some clue to the gluon spin content.
Such being the situation, it is desirable to study
the possibility that
the orbital angular momentum may contribute to some extent to the proton spin
and the orbital motion may induce intrinsic deformations of the baryons.
Non-spherical components in baryons have been discussed
in the context of the color magnetic interaction\cite{ik}, 
the electromagnetic transition\cite{ww}
and the generalized parton distribution\cite{bjy}.

 Deformed baryon states are investigated also in soliton models\cite{ws,br,lm}.
Since the topological soliton is quantized to the fixed (iso-)spin
states by rotation, the intrinsic deformations directly correlate with 
orbital motions in this description. 
Considering that the shape of the charge density is probed via the photon,
we study the Skyrme model coupled with the electromagnetic field.
Note that there are two kinds of terms which are brought about 
by the coupling of the gauge field;
non-anomalous terms through covariant derivatives
and anomalous terms through the gauged Wess-Zumino (WZ) term.

 The gauge fields 
are minimally incorporated through the covariant derivatives.
Using variation of an action obtained by the minimal replacement,
the authors of Ref.\ \cite{pt} computed
configurations of the gauge field as well as the soliton 
and estimated the magnetic moment.
They found that the dipole magnetic field is generated around the soliton
and the magnetic moment has reasonable value.
However the topological current is not taken into account  
in their variational equations.

The topological baryon current is naturally incorporated
in the Maxwell equation if one considers the gauged WZ term.
 The gauge fields affect the system not only
through the covariant derivatives but also through the gauged WZ term,
which is not obtained by the minimal replacement of derivatives
in the WZ term.
The gauged WZ term is designed to account for
the non-conservation of the axial current by the anomaly and
to describe correctly anomalous processes like
$\pi^0 \rightarrow 2\gamma$ in the pion sector.
This term possibly influences the system also in 
the soliton sector.   
Above all, the gauged WZ term is indispensable to
assure the gauge invariance of conserved currents like 
the baryon current and the electric current as well.
Actually, the gauged WZ term provides 
an anomalous contribution\cite{bpr}
of the magnetic field to the isospin charge of the soliton.

 Although the electromagnetic field is usually treated  
perturbatively owing to the small coupling constant,
it is worth estimating the anomalous contribution of the gauge field 
as the WZ term is essential for the anomalous baryon current.
 It is possible that the gauged WZ term
exerts significant influences on the physical quantities of 
the isospin fixed solitons.

In this work, we adopt an action including the gauged WZ term 
and construct spin polarized Skyrmions with a proper electric charge 
imposing boundary conditions upon the chiral field and the gauge field. 
We calculate the configuration of these fields and 
study the properties of the spin polarized Skyrmions, particularly
characteristic spatial structure of the electric charge, current
and the magnetic field.
Through these results, we discuss the intrinsic deformations 
of the charge density as well as
the significance of the effects of the electromagnetic
fields from the gauged WZ term in the soliton model.

\section{The Skyrme model with the gauged Wess-Zumino term}

 We present a model written
in terms of the chiral field and the gauge fields.
 To make it manifest how the gauge fields are incorporated in the 
system, we consider a chirally gauged model first in a general context 
and then restrict ourselves to a model coupled with 
the electromagnetic field only.
 The action of the chiral field consists of the non-linear sigma term,
the Skyrme term with pion mass term and the Wess-Zumino term as
\begin{align*}
 {\cal S}  = &   \int\! d^4x\left(\frac{F_{\pi}^2}{16}{\rm Tr}(D_\mu U 
   D^\mu U^{\dagger}) 
    +\frac{1}{32 \epsilon^2}{\rm Tr}[D_\mu U U^{\dagger},
          D_\nu U U^{\dagger}]^2 
+\frac{1}{8}m_\pi^2F_\pi^2({\rm Tr}U-2) \right)
+ \varGamma_{\rm WZ} \ .
\end{align*}
The Wess-Zumino term, $\varGamma_{\rm WZ}$, is expressed 
compactly by the use of differential forms as  
\begin{align*}
      \varGamma_{\rm WZ}=& \frac{iN_{\rm c}}{240\pi^2}\int
           {\rm Tr} ({U}^\dag {d} {U})^5   \\ &
      +\frac{iN_{\rm c}}{48\pi^2}\int {\rm Tr} \left(
        ({U}^\dag {d} {U})^3A_{\rm L}
             -({U}^\dag {d} {U})(A_{\rm L}dA_{\rm L}+
             dA_{\rm L}A_{\rm L}+A_{\rm L}^3)
        -\frac{1}{2}({U}^\dag {d} {U}A_{\rm L})^2
       + ({U}^\dag {d} {U})^2 {U}^\dag A_{\rm R} {U} A_{\rm L} 
           \right. \\
     &   
     -dA_{\rm L}  ({U}^\dag {d} {U}){U}^\dag A_{\rm R} {U}  
         -({U}^\dag {d} {U})A_{\rm L}
          {U}^\dag A_{\rm R} {U} A_{\rm L}
          \left. 
             -({U}^\dag A_{\rm R} {U})(A_{\rm L}dA_{\rm L}
             +dA_{\rm L}A_{\rm L}+A_{\rm L}^3)
        -\frac{1}{4}({U}^\dag A_{\rm R} {U}A_{\rm L})^2
             -({\rm p.c.}) 
           \right)
\end{align*}
with the abbreviation p.c.\ for parity conjugate:
 $U\leftrightarrow U^\dag, A_{\rm L} \leftrightarrow A_{\rm R}$.
The covariant derivative for the chiral field is defined as
$dx^\mu D_\mu U=dU +A_{\rm R}U -UA_{\rm L}$.

 This action except the Wess-Zumino term and the pion mass term
has the gauge symmetry and is
invariant under the gauge transformation:
\begin{align} \label{g-tr}
     U \rightarrow g_{\rm R}Ug_{\rm L}^{-1}, \
     A_{\rm L}\rightarrow g_{\rm L}(A_{\rm L}+d)g_{\rm L}^{-1}, \
     A_{\rm R}\rightarrow g_{\rm R}(A_{\rm R}+d)g_{\rm R}^{-1}.
\end{align}
 The Wess-Zumino term is constructed so as to reproduce the
chiral anomaly attended by this transformation and hence
cannot be obtained by the minimal replacement of 
differential operators which makes an action gauge-invariant. 
  The Noether current\cite{bpr} associated with the vector symmetry,
{\it i.e.} $g_{\rm L}=g_{\rm R}$, generated 
by $Q$ is
\begin{align}
 J^\mu =& -\frac{F_\pi^2}{8}
       {\rm Tr}\ iQ(UD^\mu U^\dag + ({\rm p.c}))  +\frac{1}{8\epsilon^2}
       {\rm Tr}\left([iQ,UD_\nu U^\dag][UD^\mu U^\dag,UD^\nu U^\dag]
    + ({\rm p.c.}) \right) \nonumber \\ & -
 \frac{iN_{\rm c}}{48\pi^2} \epsilon^{\mu\nu\lambda\rho}
   {\rm Tr}\ iQ [(UDU^\dag)_{\nu\lambda\rho}^3-
    \{U D_\nu U^\dag,F_{{\rm R}\lambda \rho}+ \frac{1}{2}
     U F_{{\rm L}\lambda\rho}U^\dag  \} - ({\rm p.c.})]  , \label{cur}
\end{align}
where $F_{\rm R}$, $F_{\rm L}$ are  
the field strengths for the gauge 
fields $A_{\rm R}$, $A_{\rm L}$ respectively.  
 If we set $Q=1/N_{\rm c}$ to get the baryon current,
 only the last term ---  the contribution from the WZ term ---
survives and we find the baryon number
is equivalent to the winding number. 
Therefore we can identify a baryon with a topological 
soliton in this model.

 Now we restrict ourselves to study of the SU$(2)$ chiral field
and the U$_{\rm EM}(1)$ gauge field, 
 $A_{\rm L}=A_{\rm R}= ie Q_{\rm EM} A_\mu dx^\mu$.
The electric charge matrix satisfies the Gell-Mann--Nishijima
relation: $Q_{\rm EM}=(\tau^3+1/N_{\rm c})/2$.
A soliton of a unit baryon number is realized 
in the hedgehog form : 
 $U= \exp(i{\boldsymbol \tau}\cdot\hat{\boldsymbol r} F(r))$
 with the boundary conditions
$F(r=0)=\pi, F(r=\infty)=0$.
 The electric potential is assumed to be spherical
\[eA^0 =  V(r)\]
 for simplicity, and the angular dependence
 of the vector potential is fixed 
as 
\[e\boldsymbol{A}=h(r)\sin^2\theta{\boldsymbol \nabla}\phi\] 
to match the Amp\`ere law. Here, $\theta$ and $\phi$ are
the polar and azimuthal angle respectively. 
All these do not depend on time and such a static configuration
is always taken by the gauge transformation (\ref{g-tr}).
Even a (time-dependent) collective rotation of the chiral field 
can be described by taking $g_{\rm R}=g_{\rm L}$ as the rotation matrix,
if one lets $e A^0$ take a constant value 
in proportion to the angular velocity.

 The total Lagrangian density including the kinetic term of 
the electromagnetic field is written in terms of
$F(r), V(r)$ and $h(r)$ as
\begin{align*}
  {\cal L}=& \frac{F_\pi^2}{8}\left[S_F^2V^2 \sin^2\theta -F^{\prime 2}
 -S_F^2\frac{1+(1+h\sin^2\theta)^2}{r^2}\right] 
 +\frac{S_F^2}{2\epsilon^2}\left[\left(F^{\prime 2}+\frac{S_F^2}{r^2}\right)
\right(V^2\sin^2\theta -\frac{(1+h\sin^2\theta)^2}{r^2}\left) 
-\frac{F^{\prime 2}}{r^2} \right] \\& -\frac{1}{4}m_\pi^2F_\pi^2(1-C_F) 
 +\frac{1}{4\pi^2r^2}\left[S_F^2 F^{\prime 2}V
+\frac{1}{2}S_FC_FhV'\sin^2\theta -V\left(F'h\cos^2\theta+
\frac{1}{2}C_FS_F h'\sin^2\theta\right)\right] \\
&+\frac{1}{2e^2}\left(V^{\prime 2}-4\frac{h^2}{r^4}\cos^2\theta
-\frac{h^{\prime 2}}{r^2}\sin^2\theta\right)  \ ,
\end{align*}
where $S_F=\sin F(r)$ and $C_F=\cos F(r)$. 
After the angular integration and scaling $r,V,m_\pi$ by $\epsilon F_\pi$
to be dimensionless, we are led to the variational equations;
\begin{align}
\label{eqF}
&F''+\frac{2}{r}F'+\frac{2}{r^2}
 \ (4S_F^2 F''-S_FC_F+4S_FC_ F F^{\prime 2})\left(1+\frac{2}{3}h+\frac{4}{15}h^2\right)
 -\frac{8}{r^4}S_F^3C_F\left(1+\frac{4}{3}h+\frac{8}{15}h^2\right)\nonumber \\
& \ \ +\frac{16S_F^2}{3r^2}h'F'\left(1+\frac{4}{5}h\right)-m_\pi^2 S_F
+\frac{2}{3}S_FC_F V^2\left(1-4F^{\prime 2}+8\frac{S_F^2}{r^2}\right)
-\frac{8}{3}S_F^2 V \left(F''V+2F'V'+\frac{2}{r}F'V\right) \\
&\hspace*{8em}  -\frac{\epsilon^2}{\pi^2r^2}\left(S_F^2V'-\frac{2}{3}C_F^2 hV'-
\frac{2}{3}S_F^2h'V\right)=0 \, ,  \nonumber
\end{align}
the Gauss law $\nabla\cdot {\boldsymbol E}=j^0_{\rm EM}$ with the
angular averaged charge density,
\begin{align} \label{eqV}
V''+\frac{2}{r}V'=\alpha\left[\frac{1}{\pi r^2}S_F^2F'
-\frac{2}{3\pi r^2}(C_F^2F'h+S_FC_Fh')+\frac{2\pi}{3\epsilon^2}S_F^2 V
\left(1+4F^{\prime 2}+4\frac{S_F^2}{r^2}\right)\right],
\end{align}
and an equation deduced from the Amp\`{e}re law
$\nabla \times {\boldsymbol B}={\boldsymbol j}_{\rm EM} $,
\begin{align} \label{eqh}
  h''-\frac{2}{r^2}h=\alpha\left[\frac{1}{\pi}(S_F^2F'V-S_FC_FV')+
\frac{\pi}{\epsilon^2}S_F^2\left(1+4F^{\prime 2}+4\frac{S_F^2}{r^2}\right)
\left(1+\frac{4}{5}h\right)
\right],
\end{align}
with the fine structure constant $\alpha=e^2/4\pi$.

 Note that the magnetic field ${\boldsymbol B}$
(or $h$ in r.h.s of Eq.~(\ref{eqV})) is instrumental in
producing partly the electric charge $j^0_{\rm EM}$
as a source of the electric field ${\boldsymbol E}$ 
and that ${\boldsymbol E}$  contributes the electric 
current ${\boldsymbol j}_{\rm EM}$ as a source of ${\boldsymbol B}$.
This complementarity comes from the gauged WZ term in the action
which provides the anomalous couplings between electromagnetic field and 
the chiral field like ${\boldsymbol E}\cdot {\boldsymbol B}\  \pi^0/F_\pi$.
The term brings to $j_{\rm EM}^\mu$
an anomalous current of the dual field strength as 
$\epsilon^{\mu\nu\lambda\rho}\partial_\lambda A_\rho\partial_\nu\pi^0/F_\pi$
with the non-vanishing pion field 
$\partial_\nu\pi^0/F_\pi\sim i\ {\rm Tr}\ \tau^3
(U\partial_\nu U^\dag-U^\dag\partial_\nu U)$ in the soliton sector.
In addition to this, here we emphasize that the topological term
of the baryon density is correctly incorporated in the charge density 
due to the gauged WZ term,
as one can see the first term in r.\ h.\ s.\ of Eq.\ (\ref{eqV}).

We solve the coupled equations (\ref{eqF})-(\ref{eqh})
imposing boundary conditions for the fields,
\begin{align} \label{bc}
& F(0)=\pi,& &F(r\to\infty)\propto \frac{1+\mu_\pi r}{r^2}\exp(-\mu_\pi r),
\nonumber\\
&  V'(0)=0,& &V(r\to\infty)= Z\frac{\alpha}{r} +  V_\infty{\rm(: const.)},\\
& h(r\to0)\propto r^2,& &h(r\to\infty)\propto \frac{1}{r},\nonumber
\end{align}
where $Z$ is the charge number of the soliton (to be set at 1 or 0) and 
$\mu_\pi^2=m_\pi^2-2 V_\infty^2/3$ is a modified pion mass\cite{br,dhm}
caused by `rotation' of the soliton. The angular velocity is turned to an
asymptotic constant $V_\infty$ by a gauge transformation as 
mentioned previously and thus non-zero $V_\infty$ corresponds to a `rotation'
of the soliton. In practical computing, we impose a boundary condition 
on $V'(r)$ instead of $V(r)$ without any tuning of $V_\infty$
and confirm finite values, actually.

 However, the rotation matrix does not cover the whole SU(2) group because
the generator $Q_{\rm EM}$ of U$_{\rm EM}$(1) gauge group includes 
only the $z$ component of the isospin matrix. Accordingly, $\Delta$
cannot be excluded from nucleon states without the Casimir projection 
as performed in
Ref.~\cite{anw}. In this sense, our treatment is not equivalent to 
the collective quantization of rotation fully, but to 
one-dimensional cranking. 
Nonetheless, the $z$ component of isospin (as the electric charge
minus one half of the baryon number) is assured of a quantized
value $\pm 1/2$ for the charged and neutral soliton ($Z=1,0$)
only by the boundary conditions for $F$ and $V$,
though this does not mean a topological quantization of isospin 
in contrast to that of the baryon number.
Since the asymptotic value $V_\infty$ affects the 
boundary condition of $F(r)$, the self-consistent approach
we took to solve the differential equations  
corresponds to the variation after projection.
Furthermore, the $z$-component of the spin takes $\mp 1/2$ 
for the charged and neutral solitons, since the 
spatial rotation means the inverse iso-rotation for the hedgehog configuration
and the photon contribution to the spin is negligible 
which we confirm numerically.
Therefore we regard the soliton as a spin polarized
nucleon with possible admixture of the $\Delta$ component.

\section{Field configurations and charge distribution}
\begin{figure}[t!]
  \centering
  \includegraphics[width=7.75cm,clip=yes]{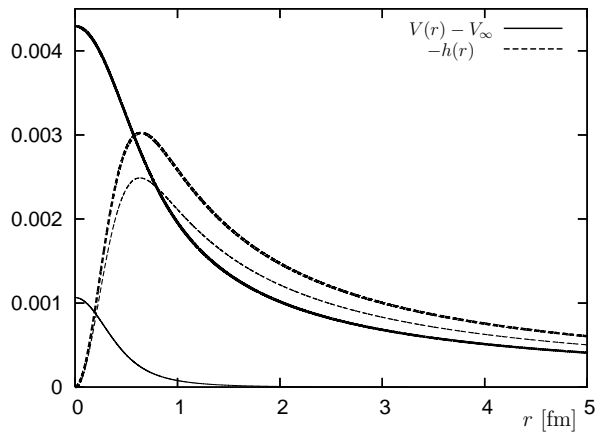}
%
  \caption{Electromagnetic fields with the parameter set II
are plotted as a function of radial distance $r$.
Bold lines are for the charged soliton and thin lines are for the neutral one.
The electric potentials (normalized by $\epsilon F_\pi$)
with the asymptotic constants subtracted 
are shown by solid lines and $-h(r)$'s which determine the magnetic field
are by dashed lines.}
  \label{fig:vh}
\end{figure}

We solve the differential equations with the boundary conditions
for proton ($Z=1$) and neutron ($Z=0$) using the relaxation method.
Three sets of parameters $F_\pi$ and $\epsilon$ are chosen 
while $m_\pi$ is fixed at the physical pion mass.
Set I: $(\epsilon=5.37,F_\pi=185\ \rm MeV)$ is determined so that
the charge radius of the proton is reproduced with physical 
pion decay constant.
Set II: $(\epsilon=5.45,F_\pi=129\ \rm MeV)$ and
Set III: $(\epsilon=4.84,F_\pi=108\ \rm MeV)$ are 
taken from Refs.\ \cite{anw} and \cite{an}, respectively.
We find that for Set I, $r_0$ such that $F(r_0)=\pi/2$ decreases 
by a few \% compared with that of Set II. 
While this means $F(r)$ for Set I is squeezed, 
$F(r)$ for Set III, on the contrary, swells compared to Set II.
Although the boundary condition constrains the fields to have
the different asymptotic behavior,
profiles of $F(r)$ with Sets II and III are almost the same as
that obtained in Refs.\ \cite{anw} and \cite{an} respectively.
The configurations of the electromagnetic field are also determined 
simultaneously, as shown in FIG.\ \ref{fig:vh}.
The electric potentials display correctly the Coulombic behavior  
for large $r$, and the smeared peak near the origin means
finite structures of soliton's charge unlike a point particle. 
The function $h(r)$ gives the magnetic field as
\begin{align} 
{e \boldsymbol B} = &\frac{h}{r^2}2\cos\theta \ \hat{\boldsymbol r}
-\frac{h'}{r}\sin\theta \ {\bf e}_\theta\ ,
\label{eB}
\end{align}
and the radial distance of their peak determines
the size of the dipole magnetic field. 
The spatial distribution of the fields discussed in detail later.
Physical quantities with these parameter sets are shown in TABLE \ref{tab}.

\begin{table}[b!]
\begin{center}
\hspace*{-1.3em}
\begin{tabular}{lc|cccccccc} 
  & & mass\hspace{.3em} &$\langle r^2 \rangle_{\rm ch}$ & 
$\langle r^2 \rangle_{\rm b}^{1/2}$ 
& $\langle r^2 \rangle_{\rm M}^{1/2}$ &  
 $\mu$ & \hspace*{1em} $g_{\rm A}$\hspace{1em}  &$g_{\pi NN}$ & $\sigma$   \\
 & & [MeV] &[fm$^2$] & [fm]&[fm]& [$\mu_N$]   &  & & [MeV]   
\\ \hline
 Set I & $p$ & 1326  & 0.757  & 0.416  & 1.128  & 3.27&
 0.616 & 6.08 &31.4 \\
 & $n$  & 1325  & -0.556  & 0.415  & 1.164  & -2.84&
 0.616 & 6.25 &31.3 \\
 Set II & $p$ & 933  & 0.773  & 0.563  & 1.118  & 3.95&
 0.572 & 8.20 &34.1 \\
& $n$  & 931  & -0.446  & 0.563  & 1.136  & -3.28&
 0.571 & 8.35 &34.0 \\ 
 Set III & $p$ & 889  & 0.890  & 0.708  & 1.086  & 5.53&
 0.672 & 11.9 &42.8 \\
&  $n$  & 887  & -0.385  & 0.707  & 1.124  & -4.79&
 0.671 & 12.1  & 42.7 \\ \hline
 exp & $p$ & 938  & 0.757  & 0.801  & 0.81  &2.79 
&1.26&13.5& $\sim$30\\
  & $n$ & 940  & -0.116 & & & -1.91 &&&
\end{tabular}
\caption{Static properties of the nucleon described by
the soliton with several sets of parameters.
$\langle r^2 \rangle_{\rm ch}$,
$\langle r^2 \rangle_{\rm b}$ and
 $\langle r^2 \rangle_{\rm M}$ are the mean square charge,
 baryon number and magnetic radii.}
\label{tab}
\end{center}
\end{table}                                

For Sets I and III, the mass of solitons deviates from the experimental 
values, but these 
receive corrections of meson loop and higher chiral-order terms\cite{mou}.
Mass differences between proton and neutron are of proper magnitude for
all sets of parameters but show the opposite sign. 
This is because the spherical ansats for the electric field is oversimple 
to give the correct sign, though we take the electric field into account 
even for the neutron.
In fact, the energy of the electric field contributes about 0.2\,\%
to the proton mass which is reasonable for the mass difference,
while it almost vanishes for the neutron case.

This setting is also reflected in the charge radius defined as
$\langle r^2\rangle_{\rm ch}=\int d^3x\  r^2\ j_{\rm EM}^0/e$ with
the charge density,
\begin{align} 
j_{\rm EM}^0 = &\frac{e}{4\pi^2 r^2}\left(-S_F^2F'-S_F^2F'h\sin^2\theta
+S_FC_Fh'\sin^2\theta +2F'h\cos^2\theta\right)  
-\frac{e}{4\epsilon^2}
\left(1+4F^{\prime 2}+4\frac{S_F^2}{r^2}\right)S_F^2V\sin^2\theta,
\label{j0}
\end{align}
which is obtained with assignment $Q=Q_{\rm EM}$ in Eq.\ (\ref{cur})
and scaling to the dimensionless variables.
For the proton, the positive charge density is piled up and the charge radius
is well reproduced comparatively. For the neutron, however, the positive
and negative charge densities compete with each other and the charge 
radius of neutron is overestimated several times as much.  

\begin{figure}[b!]
\centering
  \includegraphics[width=7cm,height=7cm]{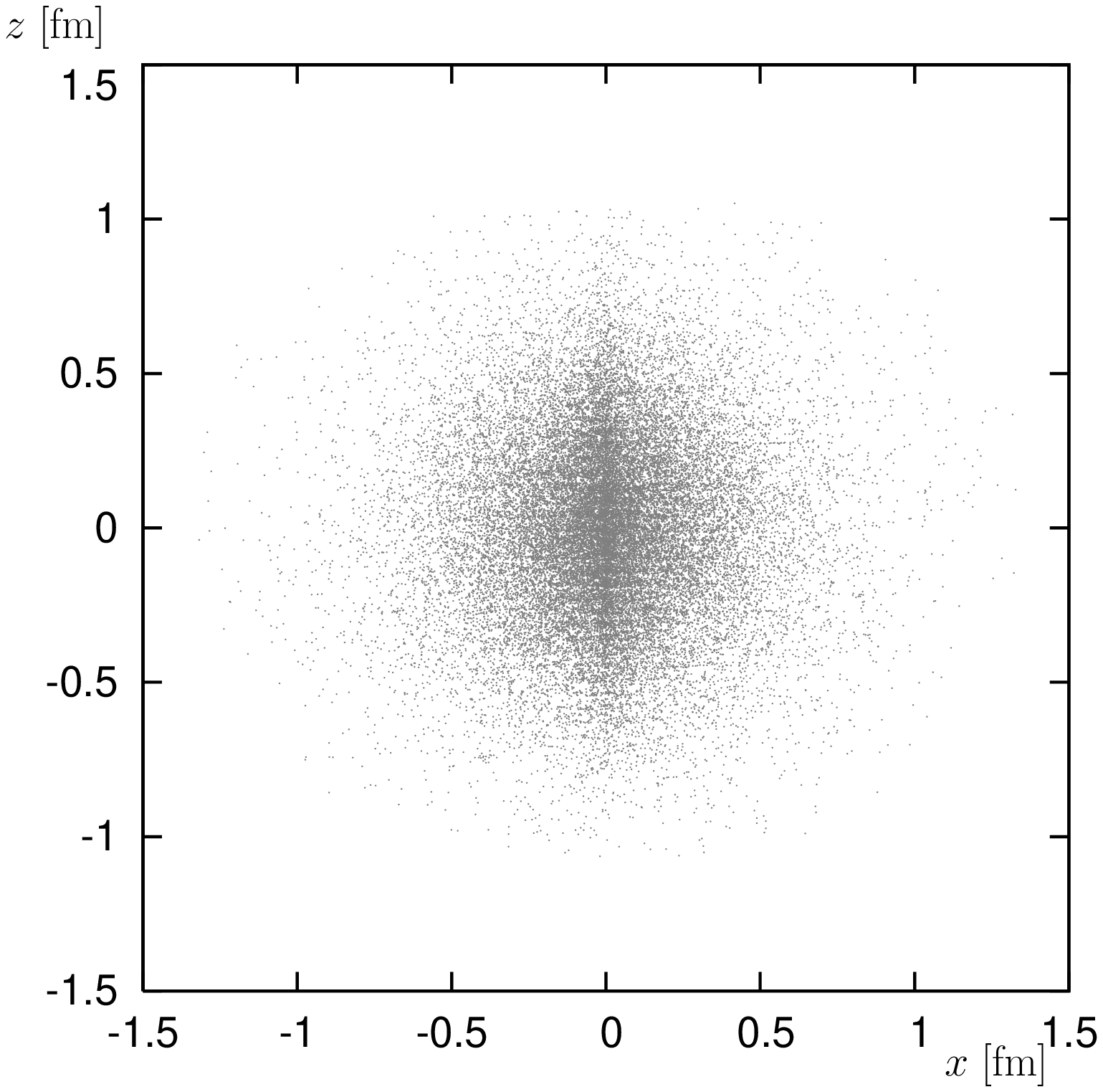} \hspace{4em}
  \includegraphics[width=7.5cm,height=7cm]{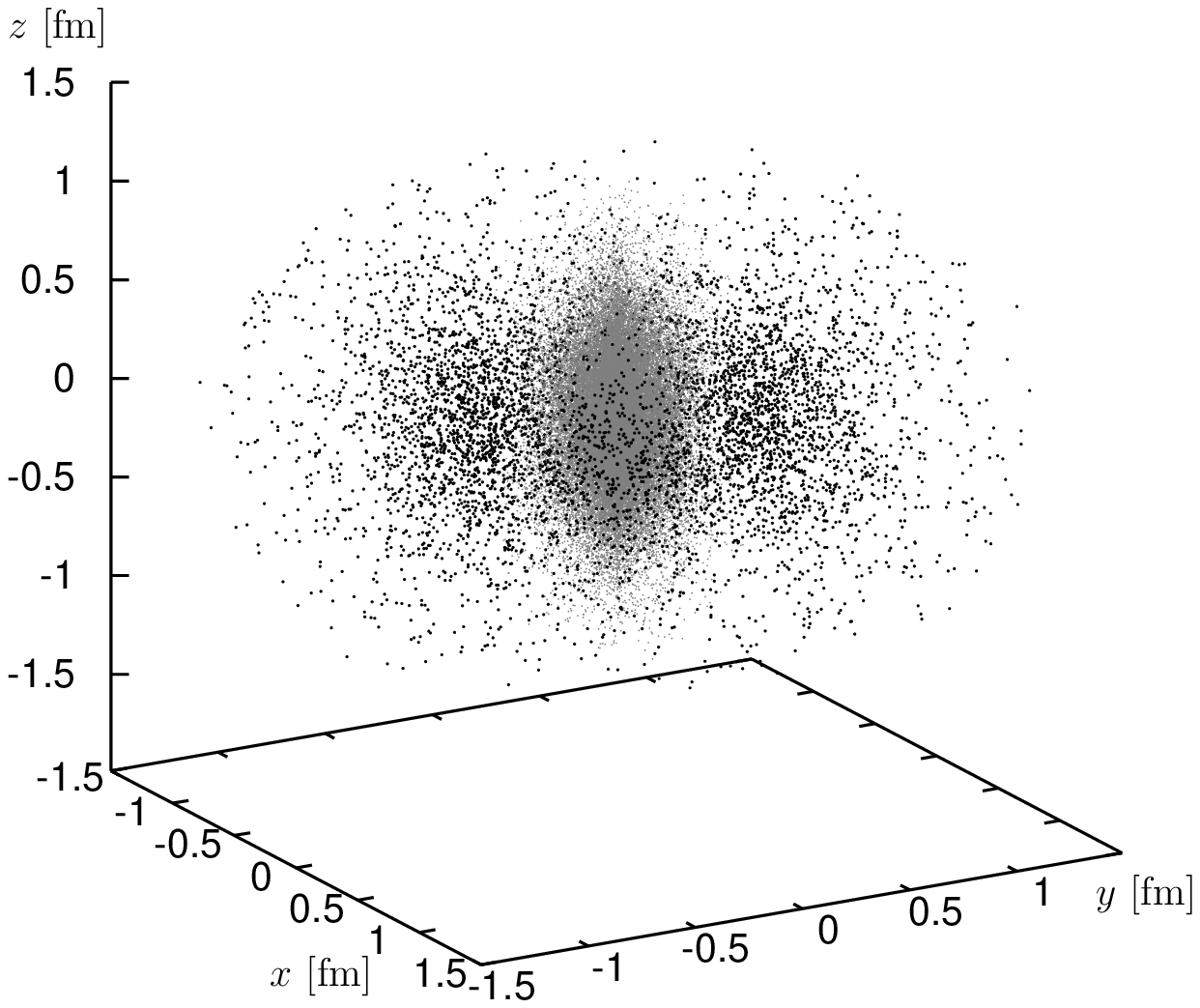}
  \caption{The charge distribution of the 
charged soliton in the $xz$ plane (left)
 and the neutral soliton in the $xyz$ space (right) calculated 
with the parameter Set II. Note that the soliton spin is polarized along
the $z$ axis.
Gray and black dots are positive and negative 
charges. The charge density of the charged soliton 
is deformed to an oblate shape and 
that of the neutral soliton has a Saturnian structure.}
  \label{fig:ch}
\end{figure}

The main contribution to the charge density, (\ref{j0}),
comes from the first term of the baryon-number density near the origin
and from the last term of the isospin charge for the medium range.
Since the latter has a remarkable polar-angle dependence,
the spatial distribution of the electric-charge density
presents a deformed or a toric structure with a core 
according to the isospin, as shown in FIG.\ \ref{fig:ch}. 

\begin{figure}[tbhp]
\begin{tabular}{rl}
\begin{minipage}{.45\hsize}
\includegraphics[width=8cm,clip=yes]{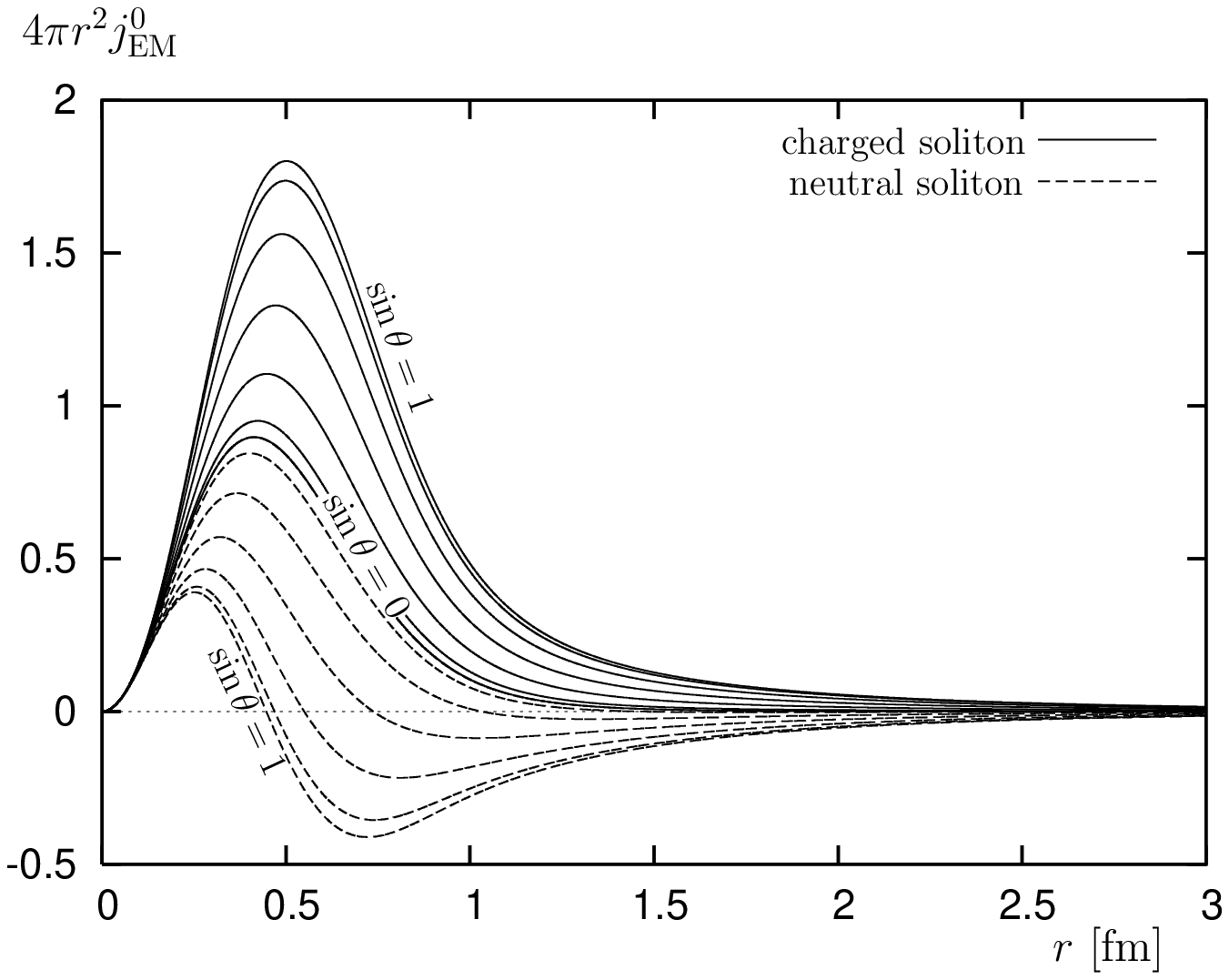}
  \caption{The charge densities as functions of the radial distance $r$
with several polar angles $\theta$.}
  \label{fig:chzen}
\end{minipage}
\hspace{2em}
\begin{minipage}{.45\hsize}
  \centering
  \includegraphics[width=8cm,clip=yes]{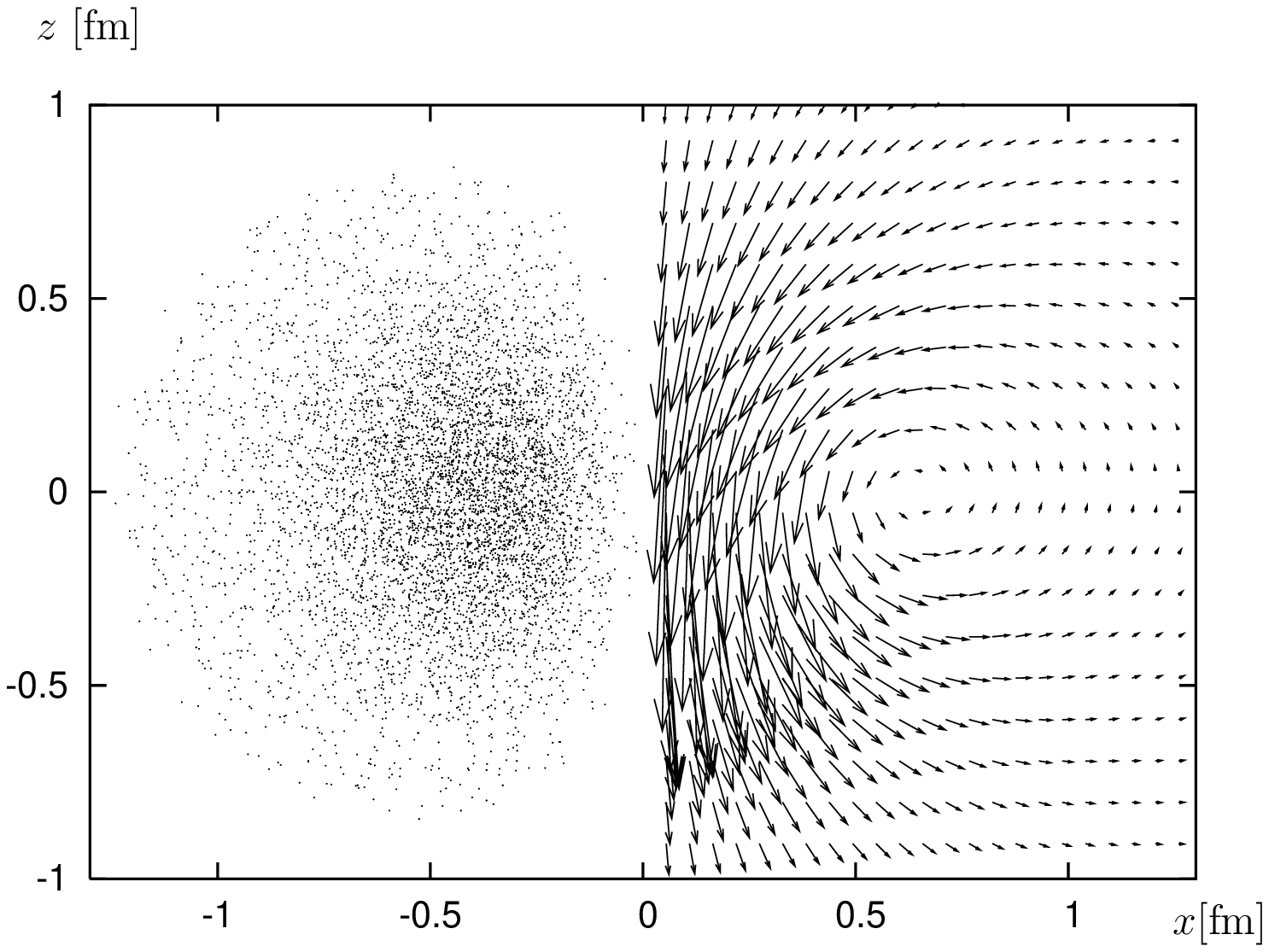}
  \caption{ The arrows in the $xz$ plane show 
the magnetic field, which is axial symmetric.
Dots in the negative $x$ region
represent the current density running around the $z$ axis clockwise.}
  \label{fig:Bj}
\end{minipage}
\end{tabular}
\end{figure}

The baryon number density depends also on the polar angle 
through the gauge field,
\begin{align*}
 b^0 =& \frac{1}{2\pi^2r^2}\left(-S_F^2F'
-S_F^2F'h \sin^2\theta + \frac{S_FC_F}{2}h'\sin^2\theta
+F'h\cos^2\theta\right)\ , 
\end{align*}
but it is almost spherical because the small fine structure 
constant suppresses the magnitude of field $h$. 
On the other hand, the isospin charge is mainly contributed
from the non-linear sigma term proportional to $S_F^2 V \sin^2\theta$
as is seen in Eq.\ (\ref{j0}).
This term affects the electric charge with the
opposite sign for charged and neutral solitons, especially
in the region such that $F(r)\sim \pi/2$ and $\sin\theta\sim 1$.
As seen in FIG.\ \ref{fig:chzen}, the solid curves of the charged soliton
change according to the polar angle and smoothly merge 
into the dashed lines of the neutral one at the polar region.
Although the angular-averaged density 
of the electric charge agrees with that obtained in Ref.\ \cite{anw},
the $\theta$ dependence makes an oblate shape for the charged soliton
and a prolate positive-charge surrounded by a toric negative-charge
for the neutral soliton.  The deformed charge density
is interpreted as a consequence of
the centrifugal force accompanying the rotation of soliton.
The radius of the negative-charge torus 
is about $0.7$ fm with the parameter Set II,
which is governed by the distance $r_0$ such that $F(r_0)\sim \pi/2$.
 Since such a deformed structure cannot be realized for the nucleon
with spin $1/2$, the density distribution is interpreted to 
stand for an intrinsic deformation of the nucleon
and also to reflect the admixed $\Delta$ component.
\hspace*{-2em}

The magnetic moment $\mu$ is evaluated from the electric current as
$\frac{1}{2}\int d^3x \ {\boldsymbol r}\times {\boldsymbol j}_{\rm EM}$
with
\begin{align}
{\boldsymbol j}_{\rm EM}= &\frac{e\sin\theta \ {\bf e_\phi} }{r}
\left[-\frac{1}{4\pi^2}S_F^2F'V+\frac{1}{4\pi^2}S_FC_FV'
-\frac{S_F^2}{4\epsilon^2}\left(1+4F^{\prime
    2}+4\frac{S_F^2}{r^2}\right)
(1+h\sin^2\theta)\right]\ ,
\label{ji}
\end{align}
which is equivalent to the estimation from asymptotic behavior
of the magnetic field, $\lim_{r\to\infty} r h(r)/\alpha$,
owing to the Amp\`{e}re law. 
Our results in unit of the nuclear magneton are comparable to the 
experimental values including their sign. 

The reasonable estimation of the magnetic moment results from
the dipole magnetic field generated by the current of soliton.
Actually, the magnetic field expressed by Eq.\ (\ref{eB})
has a dipole structure induced by the circular current Eq.\ (\ref{ji}),
as shown in FIG.\ \ref{fig:Bj}.
These field configurations are the same for both of the 
charged and neutral cases,  
while the spin of the neutral (charged) soliton is polarized (anti-)parallel
to the $z$ axis. 
The radius of the circular current is also controlled by 
the scale $r_0$ satisfying $F(r_0)\sim \pi/2$.
The spatial distribution of the magnetic field is understood 
also on an analogy with the Meissner effect in the super conductor.
The coefficient of the vector potential in the current, 
 $\left(\frac{F_\pi^2}{4}-\frac{1}{\epsilon^2} \{(\partial_\mu F)^2+\sin^2F
       (\partial_\mu \theta)^2\}\right)\sin^2F\sin^2 \theta $, 
plays a role of a square `mass' of the gauge field,
because this factor can be read
as the coefficient of $(eA_\mu)^2$ in the action.
In case of the hedgehog configuration, 
the large `mass' region forms a torus whose core,
as one can see from the factor $\sin^2 F \sin^2 \theta$,
is located at the circle around the $z$-axis such 
that $F(r)=\pi/2$ and $\theta=\pi/2$.
The configuration of the magnetic field is determined so as to
avoid the large `Meissner mass' region to save the total energy.
 As a consequence of this fact, the magnetic field
 coils around a genus of the torus.

The axial coupling tends to be one half of experiments, which 
agrees with Refs.\ \cite{anw,an}. The pion-nucleon sigma term evaluated from
the symmetry breaking term results in the adequate value as shown in 
TABLE \ref{tab}. 

In any case, contributions of the electromagnetic field via the gauged
Wess-Zumino term are controlled by the fine structure constant and
hence amount to 1\,\% of the physical quantities, at most.
Therefore the complementary dynamics of the electromagnetic field
is not conspicuous in the present situation. 
However, this anomalous contribution
might be influential in high dense matter.

\section{Concluding remarks}
 We have studied the effects of the electromagnetic field coupling
to the soliton configuration via the gauged Wess-Zumino term.
We pointed out that the gauged Wess-Zumino term provides
the Maxwell equation with anomalous terms in the presence 
of the soliton. Unless this term is considered,
the topological baryon current cannot correctly
be incorporated in the charge density of variational equations.
Furthermore non-vanishing pion fields make the anomalous coupling 
$\pi^0 {\boldsymbol E}\cdot{\boldsymbol B}$ effective in the soliton
sector, and the electric and magnetic fields contribute complementarily 
to the source current of each other. Such a current from the 
dual field strength is interesting theoretically, but in reality
these effects are suppressed by the small coupling constant.

 Because the anomalous contribution of the electric field is 
small, the spatial distribution of the electric current is mainly
determined by the non-anomalous term of the isospin current.
 From this, it is concluded that 
the electric current 
has the toric configuration,
and accordingly that the magnetic field has a poloidal structure 
coiling around the circular current according to the Amp\`{e}re law.
This dipole magnetic field is understood as a consequence of
the Meissner-like effect.
Actually, the Meissner `mass' of the gauge field is estimated 
from the chiral angle $F$ as $\sin^2 F\sin^2\theta$
corresponding to the condensate of the Cooper pairs. Contrast to the 
superconductor, the large mass region does not extend 
uniformly but forms a finite torus,
and hence the magnetic flux quantization cannot be achieved
in our case.  

 As discussed above, the symmetric axis of the torus
of the field configuration is to be interpreted physically as 
a spin-quantization axis of the soliton. This leads us
to a reasonable consequence that the magnetic moments evaluated
from the asymptotic form of the poloidal magnetic fields are  
(anti-)aligned with the spin of the charged (neutral) soliton. 
The theoretical value of the magnetic
moment derived from the variational solution is of the same order
as the experimental value of the nucleon.

We have found that
the electric charge distributions are distinctive for their polar angle
dependence. The equatorial swelling of the isospin charge makes 
the difference while the spherical core of baryon number density 
is common to charged and neutral solitons. Our results show that
the charged soliton has the oblate shape of the charge density
and the neutral soliton has the toric negative charge surrounding
the prolate shape of positive charge.
These distributions are consistent with the pion cloud
rounding the nucleon or may imply a u-d diquark core accompanied by 
the other valence quark. 
In any case, the characteristic distributions of the charge density
suggest the intrinsic deformation of baryons and 
the isospin-dependent quadrupole moment, which might be observed in
the $N\Delta$ transition.

 As in the dynamo theory for the planetary magnetic field,
a poloidal magnetic field may induce a toroidal magnetic field through
the rotation of the soliton, and {\it vice versa},
even though the physical scale is quite different. 
 It is our future work to take account of this effect by
extending the variational space of the magnetic field.
The non-radial electric field is also 
important to estimate the mass of the proton and the neutron
and to discuss a correlation between the orbital motion
and the intrinsic deformation.

\section*{Acknowledgments}
One of the authors (M.~O)
 would like to acknowledge enlightening discussions with K.~Yazaki.
He is indebted to O. Morimatsu, H. Fujii and H. En'yo for useful comments.
This work was supported in part by the Japan Society for the Promotion
of Science for Young Scientists.

\end{document}